\documentstyle{article}\begin{document}
\title{Non-linear relativistic diffusions}
\author{ Z. Haba\\
Institute of Theoretical Physics, University of Wroclaw,\\ 50-204
Wroclaw, Plac Maxa Borna 9, Poland\\
email:zhab@ift.uni.wroc.pl}\maketitle
\begin{abstract} We obtain a non-linear generalization of
the relativistic diffusion.  We discuss diffusion equations whose
non-linearity is a consequence of quantum statistics.We show that
the assumptions of the relativistic invariance and an interpretation
of the solution as a probability distribution substantially restrict
the class of admissible non-linear diffusion equations. We consider
relativistic invariant as well as covariant frame-dependent
diffusion equations with a drift. In the latter case we show that
there can exist stationary solutions of the diffusion equation
besides the equilibrium solution corresponding to the quantum or
Tsallis distributions. We define the relative entropy  as a function
of the diffusion probability and prove that it is monotonically
decreasing in time when the diffusion tends to the equilibrium. We
discuss its relation to the thermodynamic behavior of diffusing
particles .\end{abstract}
 \section{Introduction}
An interaction of a particle with a many particle system is too
complex to be described by a microscopic Hamiltonian theory. Some
approximations for the interaction are unavoidable. If the medium is
passive then an approximation of a linear master equation is
sufficient \cite{pitaj}. In the quantum case there is a restriction
on such an approximation (even if the interaction is weak) resulting
from quantum statistics. The probability of a transition to a
quantum state depends on the occupation number of this state.  Such
a dependence implies a non-linear master equation  \cite{2}. The
collision terms in the Boltzmann equation \cite{groot} as well as
interactions in an equation for the Wigner function \cite{zachar} in
quantum field theory
 also lead to non-linear integro-differential equations.
Under the
 approximation of two-body interactions the non-linearities are quadratic.
 The
approximation of binary collisions together with a non-relativistic
approximation to the probability distribution of the electron gas is
applied to electron-photon interactions \cite{komp}\cite{rybicki}
leading to the Kompaneets diffusion equation for the photon
distribution. The simplicity of the Kompaneets equation  inspired us
for a search of relativistic partial differential equations of the
parabolic type.

In this paper we discuss diffusion equations from the point of view
of the general principles of relativistic invariance. First, we
consider relativistic invariant non-linear diffusion equations in
the phase space $(x,p)$.  In general, differential equations
invariant under a group of transformations $L$ have solutions which
transform in a non-trivial way under $L$. We consider rotation
invariant stationary solutions. They transform covariantly under the
Lorentz group. After a transformation the solution depends on the
velocity $w^{\mu}$ of the reference frame (the boost). We may treat
the frame velocity as an additional physical variable and discuss
relativistic equations for densities which depend on the phase space
variables $(x,p)$ and on $w$. In such a case we obtain a wider class
of non-linear diffusion equations similar to the Kompaneets
equation.

The linear relativistic diffusions have been studied for a long
time (see the reviews \cite{debrev}\cite{hang}). It has been shown
a long time ago that Markovian relativistic diffusion in the
configuration space does not exist \cite{lop}\cite{hakim}. The
relativistic diffusion (preserving the mass-shell) of a massive
particle on the phase space has been defined and studied by Schay
\cite{schay} and Dudley \cite{dudley}. A general class of
diffusions in the phase space has been discussed in
\cite{chevalier}. An analog of Dudley's diffusion in general
relativity has been investigated in \cite{lejan}. We have studied
diffusions with a friction of massive and massless particles in
Minkowski space \cite{haba} and on a general manifold
\cite{habacqg}. These diffusions have an equilibrium limit
determined by the friction. Linear diffusion equations are applied
in the quark-gluon plasma and in heavy ion collisions
\cite{hwa}\cite{svet}\cite{rafelski}\cite{ion} (for non-linear
diffusions see \cite{wol}). The relativistic invariance in these
applications is not clear because only diffusion of some momentum
variables is investigated (e.g. diffusion of the rapidity). The
first diffusion equation applied to relativistic phenomena
appeared in the paper of Kompaneets \cite{komp} (although the
electron gas was treated in a non-relativistic approximation and
only the diffusion of the photon energy was considered).
Non-relativistic non-linear diffusion equations appear in porous
media,plasma physics, stellar dynamics  and most importantly in
hydrodynamics \cite{zeld2}. A non-linear equation in the phase
space (Kramers equation) can be related to a non-linear equation
in the configuration space in the high friction limit
\cite{bertini}. The statistical physics of non-linear diffusion
equations has been discussed recently in \cite{ott}\cite{hav}. It
seems that power-like non-linearities in the diffusion equation
are intrinsically connected with the Tsallis statistics
\cite{tsallisnon}\cite{plastino}\cite{tsallis}\cite{nonext}\cite{book}.

The organization of this paper is as follows. In sec.2 we begin
with relativistic quantum mechanics in order to discuss the  form
of the nonlinear master equation. We treat  St\"uckelberg's proper
time formalism in relativistic quantum mechanics
\cite{stuck}\cite{feyn}\cite{hor} as a useful heuristic tool for a
derivation of relativistic invariant equations. We consider the
relativistic diffusion as a classical limit of the master
equation. We propose a non-linear Lorentz covariant relativistic
diffusion equation preserving the positivity and normalization of
the probability distribution.  In sec.3 we recall the model of the
linear relativistic diffusion (we need this simple model to fix
the notation). In sec.4 we obtain a non-linear diffusive
relativistic transport  equation. We begin the studies of its
consequences. First, we show that the transport equation has a
stationary solution but no equilibrium. In sec.5
 relativistic covariant drifts are discussed. It is pointed out
 that an additional tensor is needed in order to define the drift.
 A unit four-vector $w^{\mu}$ which can be related to the frame velocity is introduced
 in order to construct the drift. The resulting diffusion equation
 has an equilibrium. A dependence of the physical equilibrium state on the
 relativistic frame of reference has been discussed in physics literature for a long time
 (see \cite{temp} and references quoted there).
  We consider quadratic and power-like non-linearities.
 In sec.6 we show that the linear approximation coincides with the
 diffusion  discussed in \cite{haba}
 and \cite{habampa}(equilibrating to the J\"uttner distribution, see also earlier papers
 cited in \cite{debrev}\cite{hang}). In sec.7 we define the relative
 entropy for the non-linear diffusions.
 We discuss drifts leading to the quantum equilibrium
 or to the Tsallis distribution. We show that the relative entropy is decreasing
 monotonically when the diffusion tends to its equilibrium.

\section{ Irreducible
representations of the Poincare group and  relativistic diffusion
equations} Relativistic invariance in quantum physics should be
treated by means of the representation theory of the Poincare
group (irreducible representations if elementary particles are
discussed). Let $A\in SL(2,C)$ and $\Lambda$ be a homomorphism of
$SL(2,C)$ onto $SO(3,1)$. We consider one-particle states
(described by a function of the momentum) transforming under an
irreducible unitary representation $U$ of the Poincare group
\cite{wein}\cite{ohnuki}. When restricted to the Lorentz group the
representation is expressed on eigenvectors $\vert {\bf
p},\sigma\rangle$ of the momentum ${\bf P}$ and the third
component of the spin $\Sigma_{3}$($-2j-1\leq \sigma\leq
2j+1$)\begin{equation} U(A)\vert p\rangle=V({\cal
R}(A))\vert\Lambda(A^{-1})p\rangle,
\end{equation} where $p^{\mu}p_{\mu}=m^{2}c^{2}$,  ${\cal R}$ is the Wigner
rotation and the matrix $V$ is defined by  $(2j+1)$-dimensional
representation of $SU(2)$ .  We obtain  generators of the Lorentz
group by differentiation (Greek indices denote space-time
coordinates whereas the  Latin indices refer to spatial
coordinates)
\begin{equation}
M_{\mu\nu}=L_{\mu\nu}+\Sigma_{\mu\nu}(p),
\end{equation}
where \begin{equation} L_{jk}=-i(p_{j}\frac{\partial}{\partial
p^{k}}-p_{k}\frac{\partial}{\partial p^{j}}),
\end{equation}
\begin{equation} L_{0j}=-ip_{0}\frac{\partial}{\partial
p^{j}}
\end{equation}
and $\Sigma_{\mu\nu}\in su(2)$ .

We outline here the scheme which leads to a relativistic diffusion
equation discussed in more detail in \cite{habajpa}(the linear
theory).
 For any classical observable $O$ we can define its operator Weyl version ${\cal O}$.
 Then, we consider matrix elements of ${\cal O}$
 in
 the basis of the wave functions (1)
\begin{displaymath} {\cal O}_{\sigma\sigma^{\prime}}(p,p^{\prime})
=\langle p,\sigma\vert {\cal O}\vert
p^{\prime},\sigma^{\prime}\rangle.
\end{displaymath}
We define a classical function on the phase space  (the Wigner
matrix) $W$ as the Fourier transform of ${\cal O}$ (its limit
$\hbar\rightarrow 0$ should coincide with $O$)
\begin{equation}\begin{array}{l}
W_{\sigma\sigma^{\prime}}({\bf x},{\bf p}) =\int d{\bf k}d{\bf
k}^{\prime}\int\delta({\bf p}-\frac{1}{2}{\bf k}-\frac{1}{2}{\bf
k}^{\prime})  \exp(i({\bf k}-{\bf k}^{\prime}){\bf x}){\cal
O}_{\sigma\sigma^{\prime}}(k,k^{\prime}).\end{array}
\end{equation}
In the massless case instead of $2j+1$ states of spin $j$ we have
only two states of helicity $\lambda=j$ and $\lambda=-j$.
 The formulae for
generators of an irreducible representation of the Poincare group
have been discussed in our earlier paper following
\cite{wein}\cite{moses}\cite{foldy}\cite{shirokov}\cite{birula}.

We consider relativistic quantum mechanics formulated in an
invariant way by means of the proper time $\tau$
\cite{stuck}\cite{feyn}\cite{hor}. We look for non-linear master
equations consistent with quantum mechanics. A non-linearity of
the master equation arises as a consequence of the quantum
statistics \cite{2} or in a description of particles interacting
by means of an electromagnetic field
\cite{lenz}\cite{castin}\cite{castin2}. In the first case, the
probability of a transition depends on the occupation number of
the state.  In the second case an electromagnetic field depends on
the distribution of charged particles which is described by the
density matrix $\rho$. If we express the electromagnetic field by
its source $\rho $ then we obtain a non-linear master equation. In
general, such an equation will have a complicated nonlocal form.
We consider a local approximation independent of details of
particular interactions. We assume that the operators in the
equation can be expressed solely by generators of the Poincare
group and some scalars. We demand that after a time evolution  the
density matrix remains a Hermitian operator and that its trace is
preserved. We additionally assume that the generator of the
evolution is a quadratic function of the generators of the
Poincare group (this assumption leads to the diffusion). Then, the
structure of the evolution equation can be derived in a similar
way as in the case of the Lindblad equation \cite{lindblad}. Our
assumptions suggest the equation
\begin{equation}\begin{array}{l}
\partial_{\tau}\rho=i[P^{\mu},[P_{\mu},\rho]]-\frac{\gamma^{2}}{8}
([M^{\mu\nu},[GM_{\mu\nu},\rho]]+[GM^{\mu\nu},[M_{\mu\nu},\rho]]).
\end{array}\end{equation} Here, $\gamma^{2}$ is a diffusion
constant which could be determined from the physical model of
particle interactions, $G(\rho)$ is an operator depending on
$\rho$ which is a scalar with respect to the transformations of
the Lorentz group. Eq. (6) is covariant under transformations of
the Lorentz group. If there is another tensor $F_{\mu\nu}$ at our
disposal then we may insert $M_{\mu\nu}+F_{\mu\nu}$ in eq.(6). We
shall discuss this possibility in sec.5.

In the linear case ($G=1$) the time evolution of an observable
${\cal O}$ can be defined by the time evolution of the state
$\rho$ (here the trace defines an expectation value of the
observable)
\begin{equation}
Tr(\rho_{\tau}{\cal O})=Tr(\rho {\cal O}_{\tau}).
\end{equation} From the relation (7)
between the evolution of states and the evolution of observables it
follows that an evolution equation  for observables also preserves
the positivity of an observable and its trace.
 Then, from eqs.(6) and
(7) we obtain (see \cite{habajpa} and \cite{groot}\cite{zachar}) a
linear equation for $W$ (defined in eq.(5)). We generalize this
equation ( when $G\neq 1$) to a non-linear diffusion equation
\begin{equation}\begin{array}{l}
\partial_{\tau}W=p^{\mu}\partial_{\mu}^{x}W
-\frac{1}{4}\gamma^{2}(M_{\mu\nu}GM^{\mu\nu}W
+WGM_{\mu\nu}M^{\mu\nu}\cr
-GM_{\mu\nu}WM^{\mu\nu}-M_{\mu\nu}WGM^{\mu\nu})\end{array}
\end{equation}where $G(W) $ is a Lorentz scalar but in general a non-linear function of
$W$, the operators on the rhs of $W$ act to the left (their position
is relevant only for particles with spin, see \cite{habajpa}).
\section{Linear diffusion of  particles without spin}

 Let us  write explicitly the linear version ($G=1$) of eq.(8) for particles with the spin equal zero (see
\cite{habajpa} for a higher spin)
\begin{equation}\begin{array}{l}
\kappa^{-2}(\partial_{\tau}W-p^{\mu}\partial^{x}_{\mu}W)=\frac{1}{2}\triangle_{H}^{m}W,
\end{array}
\end{equation}
where $\kappa^{2}=\gamma^{2}m^{-2}c^{-2}$ and \begin{equation}
\triangle_{H}^{m}=m^{2}c^{2}(\partial_{1}^{2}+\partial_{2}^{2}+\partial_{3}^{2})+p^{j}p^{k}\partial_{j}\partial_{k}
+3p^{k}\partial_{k}
\end{equation} $k=1,2,3$ and $\partial_{j}=\frac{\partial}{\partial
p^{j}}$. The derivatives over position will have an index $x$,
derivatives without an index are over momenta. We can take the
limit $m\rightarrow  0$ of the operator (10). Then, the limit
$m\rightarrow 0$ of eq.(9) is defined by
\begin{equation}
\triangle_{H}^{0}=p^{j}p^{k}\partial_{j}\partial_{k}
+3p^{k}\partial_{k}.
\end{equation}
We can see that the operator (10) is the Laplace-Beltrami operator
on the hyperboloid $p^{\mu}p_{\mu}=m^{2}c^{2}$
\begin{equation}
\triangle_{H}^{m}=g^{-\frac{1}{2}}\partial_{j}g^{jk}g^{\frac{1}{2}}\partial_{k},
\end{equation}where
\begin{equation}
g^{jk}=m^{2}c^{2}\delta^{jk}+p^{j}p^{k},
\end{equation}
$g=\det(g_{jk})$ and $g^{-1}=m^{6}c^{6}p_{0}^{2}$,
$p_{0}=\sqrt{m^{2}c^{2}+{\bf p}^{2}}$. Note that in terms of the
spatial components of the four-vector momentum the Lorentz
transformation ${\bf p}\rightarrow {\bf p}^{\prime}$ is
non-linear. The invariance of the metric tensor takes the form
\begin{equation}
g^{jk}({\bf p}^{\prime})=J_{jl}g^{lr}({\bf p})J_{kr},
\end{equation} where \begin{equation} J_{jl}=\frac{\partial
p^{\prime j}}{\partial p^{l}}.
\end{equation}
Using eq.(14) we can check by direct calculations that
transforming coordinates by means of the Lorentz transformation we
do not change the Laplace-Beltrami operator (what is well-known;
however, the explicit  calculations checking  the invariance of
the non-linear diffusion will be useful later on)\begin{equation}
\triangle^{m}_{H}({\bf p}^{\prime})=\triangle^{m}_{H}({\bf p}).
\end{equation}

 In the classical description we define the
evolution of the probability density $p_{0}\Phi$ by an adjoint
equation
\begin{equation}\begin{array}{l}
(\Phi_{\tau},W)\equiv\int dxd{\bf p}\Phi_{\tau}(x,{\bf p})W(x,{\bf
p})\equiv \int dxd{\bf p}\Phi(x,{\bf p})W_{\tau}(x,{\bf
p})=(\Phi,W_{\tau}). \end{array}\end{equation} We define the
adjoints of operators with respect to the (real) scalar product
(17)
\begin{equation}
(\Phi,AW)=(A^{*}\Phi,W). \end{equation} Note that
$L_{jk}^{*}=-L_{jk}$ and $L_{0j}^{*}=i\partial_{j}p_{0}$.

 We obtain the transport equation if both
sides of eq.(17) do not depend on $\tau$. This is the case if
\begin{equation}
-p^{\mu}\partial_{\mu}^{x}W=\frac{\kappa^{2}}{2}\triangle_{H}^{m}W
\end{equation}
or equivalently
\begin{equation}
p^{\mu}\partial_{\mu}^{x}\Phi=\frac{\kappa^{2}}{2}\triangle_{H}^{m
*}\Phi,
\end{equation}
where $\triangle_{H}^{m *}$ is the adjoint (18) of
$\triangle_{H}^{m } $ (this is also the adjoint in $L^{2}(d{\bf
p})$). We write $\Omega=p_{0}\Phi$. Then, eq.(20) can be rewritten
in the form\begin{equation}\begin{array}{l}
p^{\mu}\partial_{\mu}\Omega=
\frac{\kappa^{2}}{2\sqrt{g}}\partial_{j}(\sqrt{g}g^{jk}\partial_{k}\Omega).\end{array}
\end{equation}

\section{Non-linear relativistic diffusion equations (no drift)}
We restrict ourselves to particles with spin zero from now on.
Eq.(8) can be expressed by spatial components ${\bf p}$ of the
four-vector $p$
 as
\begin{equation}\begin{array}{l}
\partial_{0}\Omega-p_{0}^{-1}{\bf p}\nabla_{{\bf x}} \Omega=
\frac{1}{2}\kappa^{2}\partial_{j}\Big(G(\Omega)
g^{jk}p_{0}^{-1}\partial_{k}\Omega \Big)\end{array}
\end{equation}
or in a more elegant way
\begin{equation}\begin{array}{l}
p^{\mu}\partial_{\mu}\Omega=
\frac{\kappa^{2}}{2\sqrt{g}}\partial_{j}\Big(\sqrt{g}g^{jk}\partial_{k}{\cal
G}\Big)\end{array}
\end{equation} when \begin{equation} G(\Omega)={\cal
G}(\Omega)^{\prime}.
\end{equation}

 Eq.(22) is in a divergence form
\begin{equation}
\partial^{A}J_{A}=0,
\end{equation}
where $(x^{A})=(x,{\bf p}).$ As a consequence the particle number
$N$ is conserved ($x_{0}=ct$)\begin{equation}
\partial_{t}N=\partial_{t}\int d{\bf x}d{\bf p}\Omega=0.
 \end{equation}
We could show that the preservation of the probability (26) is a
classical equivalent of the quantum conservation laws
\begin{equation}
\partial_{t}Tr(p_{0}{\cal O}_{t})=\partial_{\tau}Tr({\cal
O}_{\tau})=0.
\end{equation}
 We are looking for solutions of eq.(22) depending only on
 dimensionless energy\begin{equation}
 \epsilon=\beta c p_{0},
 \end{equation}where $\beta^{-1}$ is a parameter of the dimension of the energy which we can
 set as equal to $(k_{B}T)^{-1}$ (where $T$ is the temperature and $k_{B}$ the Boltzmann constant).
Then,\begin{displaymath}
\partial_{j}\Omega(\epsilon)=\beta cp_{j}p_{0}^{-1}\partial_{\epsilon}
\Omega(\epsilon).
\end{displaymath}If $\Omega $ depends only on $\epsilon$ then
eq.(22) reads

\begin{equation}\begin{array}{l}
\partial_{t}\Omega_{t}(\epsilon)=\frac{\beta
c^{2}\kappa^{2}}{2}\Big(\epsilon^{-2}\partial_{\epsilon}(\epsilon^{3}-\epsilon
m^{2}c^{4}\beta^{2})\partial_{\epsilon}{\cal
G}(\Omega_{t})+m^{2}c^{4}\beta^{2}\epsilon^{-2}\partial_{\epsilon}{\cal
G}(\Omega_{t})\Big).\end{array}
\end{equation}
First, assume $m>0$, then  the stationary solution
$\partial_{0}\Omega_{S}=0$ is determined by the equation (note
that the current (25) $J\neq 0$, hence $\Omega_{S}$ is not an
equilibrium)
\begin{equation} (\epsilon^{3}-\epsilon
m^{2}c^{4}\beta^{2})\partial_{\epsilon}{\cal
G}(\Omega_{S})+m^{2}c^{4}\beta^{2}{\cal G}(\Omega_{S})=R,
\end{equation} where $R$ is a constant. If ${\cal G}$ is known then
integrating  eq.(30) we obtain an implicit equation for $\Omega_{S}$
\begin{equation}
{\cal
G}(\Omega_{S})=\frac{\epsilon}{\sqrt{\epsilon^{2}-m^{2}c^{4}\beta^{2}}}.
\end{equation}
If the stationary state $\Omega_{S}(\epsilon)$ is given
 then eq.(31) determines ${\cal G}$ as a function of $\Omega$.

In the massless case, let the initial $\Omega$ depends on ${\bf x}$
and $\beta c{\bf p}={\bf n}\epsilon$. Then, eq.(22) reads
\begin{equation}
\partial_{t}\Omega_{t}(\epsilon)-c{\bf n}\nabla_{{\bf x}}\Omega_{t}=\frac{\beta
c^{2}\kappa^{2}}{2}\epsilon^{-2}\partial_{\epsilon}\epsilon^{3}\partial_{\epsilon}{\cal
G}(\Omega_{t}).
\end{equation}
Let $\Omega_{t}({\bf x},{\bf n},\epsilon)$ be the solution of
eq.(29) (with $m=0$, ${\bf x}$ and ${\bf n}$ treated as  fixed
parameters) then the solution of eq.(32) is
 $\Omega_{t}({\bf x}-{\bf n}ct,{\bf
n},\epsilon)$ . Solutions of eq.(30) have a discontinuity at $m=0$.
The limit $m\rightarrow 0$ of eq.(31) exists  but is trivial
 (${\cal G}=const$ corresponding to $G=0$). There exist non-trivial
stationary solutions of eq.(32) (which are not a limit of the
solution (31)). In fact, in the massless case the condition
$\partial_{t}\Omega_{S}=0$ gives
\begin{equation}{\cal G}(\Omega_{S})=a^{-2}\epsilon^{-2}
\end{equation}
where $a$ is an arbitrary integration constant.

Power-like nonlinearities in non-relativistic  diffusion equations
have been discussed in \cite{tsallisnon}\cite{plastino}. A
relativistic non-linear diffusion as a function of the rapidity
has been studied in \cite{wol} and applied to the heavy ion
collisions. In our model the Tsallis distribution
\cite{book}\cite{tsallis0}
\begin{equation}
\Omega_{S}=(1+(q-1)\epsilon)^{-\frac{1}{q-1}}
\end{equation}
results as a stationary distribution if $m=0$, $a=q-1$ and

\begin{equation}
{\cal G}=(q-1)^{-2}\Omega^{2q-2}(1-\Omega^{q-1})^{-2}.
\end{equation}

From the Lorentz invariance of the diffusion equation (22) it
follows that if $\Omega_{S}(\epsilon)$ is a stationary solution of
eq.(22) then $\Omega_{S}(\omega)$ is also a stationary  solution of
this equation where
\begin{equation}
\omega=\beta c p^{\mu}w_{\mu} \end{equation} and $w_{\mu}$ is an
arbitrary unit time-like four-vector. We can show a more general
result: assume that $\Omega$ depends only on $\omega$ and on $
X=x^{\mu}w_{\mu}$ then eq.(23) can be expressed in the form

\begin{equation}\begin{array}{l}
\partial_{X}\Omega_{X}(\omega)=\frac{\beta
c^{2}\kappa^{2}}{2}\Big(\omega^{-2}\partial_{\omega}(\omega^{3}-w^{\mu}w_{\mu}\omega
m^{2}c^{4}\beta^{2})\partial_{\omega}{\cal G}(\Omega_{X})\cr
+w^{\mu}w_{\mu}m^{2}c^{4}\beta^{2}\omega^{-2}\partial_{\omega}{\cal
G}(\Omega_{X})\Big).\end{array}
\end{equation}
An $X$-independent solution of eq.(37) can be derived in the same
way as in eqs.(30)-(31) with $\epsilon\rightarrow\omega$ and
$m^{2}c^{4}\rightarrow m^{2}c^{4}w^{\mu}w_{\mu}$.We recover
eq.(29) when $w=(1,0,0,0)$.

\section{Frame-dependent drifts of non-linear diffusion equations}
In sec.2 we suggested that the non-linear relativistic diffusion
equation is essentially unique up to an arbitrary (state
dependent) diffusion strength $G$. In sec.4 we have shown that the
relativistic invariant diffusion equation has a stationary
solution $\Omega_{S}$ which is not an equilibrium ($J\neq 0$). In
this section we construct Lorentz covariant drifts such that the
resulting diffusion equation has an equilibrium (in addition to
the stationary states of sec.4). We could begin (as in sec.2) with
the representation theory of the Poincare group. The first order
operator $M_{\mu\nu}$ transforms as a tensor. In order to define a
drift we need another tensor to couple $M_{\mu\nu}$ in an
invariant way. As an example, an interaction with the
electromagnetic field $F^{\mu\nu}$ can be treated as a drift
$F^{\mu\nu}M_{\mu\nu}W$. However, if the equilibrium is to be
space-time independent then such a drift is excluded. We cannot
couple $x^{\mu}$ to $M_{\mu\nu}$ for the same reason. The tensor
$p_{\mu}p_{\nu}$ being symmetric does not couple to $M_{\mu\nu}$.
We need some other vectors (or tensors) to achieve such a
coupling. In sec.4 we have expressed the stationary state
$\Omega_{S}$ as a function of the energy. We have shown that a
Lorentz transformation of $\Omega_{S}$ depending on the velocity
$w^{\mu}$ of the Lorentz frame is again a stationary solution of
the diffusion equation. The velocity $w^{\mu}$ is determined in
the unique way. We could assume from the beginning that $w^{\mu}$
is a variable in the diffusion equation (on physical grounds the
equilibrium should depend on the Lorentz frame \cite{temp}) . In
such a case we obtain a wider class of covariant diffusion
equations. We consider diffusion equations which in the rest frame
and in a linear approximation reduce to the diffusion equations
discussed in \cite{schay}\cite{dudley}\cite{haba}(Boltzmann
statistics).
 Assuming that there is an antisymmetric tensor $A_{\mu\nu}$ available
 we suggest  the following  non-linear diffusion equation with a drift
($M_{\mu\nu}=L_{\mu\nu}$ for a particle without spin,
$L^{*\mu\nu}$ is defined below
eq.(18))\begin{equation}\begin{array}{l}
p^{\mu}\partial^{x}_{\mu}p_{0}^{-1}\Omega=\frac{1}{2}\kappa^{2}
iL^{*\mu\nu}G\Big(iL^{*}_{\mu
\nu}p_{0}^{-1}\Omega+A_{\mu\nu}(\Omega)p_{0}^{-1}\Omega\Big).
\end{array}\end{equation} In ${\bf p}$ coordinates
eq.(38) reads \begin{equation}\begin{array}{l}
\sqrt{g}p^{\mu}\partial^{x}_{\mu}\Omega=
\frac{1}{2}\kappa^{2}\partial_{j}\Big(G
g^{jk}\sqrt{g}(\partial_{k}\Omega
+A_{k\mu}p^{\mu}\Omega)\Big).\end{array}
\end{equation}
We are looking for  equilibrium solutions $\Omega_{E}$ of eq.(39)
depending on $\epsilon$. It is easy to see that such a solution is
determined by the equation (the current $J$ of eq.(25) is zero)
\begin{equation}
 g^{jk}p_{0}^{-1}\partial_{k}\Omega_{E}
+A^{0j}\Omega_{E}=0
\end{equation}
(the term $A^{jk}$ is absent in eq.(40) because of the rotation
invariance of $\Omega_{E}$ and the antisymmetry of $A^{jk}$).

As an example of relativistic covariant eqs.(38)-(39) we could
consider an interaction of a particle with an electromagnetic
field (however, this does not serve our purpose of  finding new
equilibria). In an electromagnetic field
$A_{\mu\nu}=F_{\mu\nu}=\partial_{\mu}A_{\nu}-\partial_{\nu}A_{\mu}$
 eq.(39) takes the form
\begin{equation}\begin{array}{l}
\partial_{0}\Omega-p_{0}^{-1}{\bf p}\nabla_{{\bf x}} \Omega=
\frac{1}{2}\kappa^{2}\partial_{j}\Big(G(\Omega)
g^{jk}p_{0}^{-1}\partial_{k}\Omega
+F_{k\nu}p^{\nu}p_{0}^{-1}\Omega\Big).\end{array}
\end{equation}
It can be shown that if $\Omega_{S}(\epsilon)$ is a stationary
solution of eq.(22) then the stationary solution of eq.(41) in a
scalar potential $A_{0}$  is $\Omega_{S}(\epsilon + A_{0})$.

We could also approach the construction of a relativistic
 covariant drift directly (without referring to the representation theory of the Poincare group).
  So, if $\phi$ is a scalar
 then it follows from eq.(14) that (for an arbitrary scalar $H(\Omega)$) the drift
 \begin{displaymath}
 H(\Omega)g^{jk}\partial_{j}\phi\partial_{k}\Omega
 \end{displaymath}
 is relativistic invariant.
If the equilibrium is to be independent of $x$ then we should
construct the scalar from $p$ and another vector $w^{\mu}$
($p^{2}=m^{2}c^{2}$ is  a constant). We put $\phi=w^{\mu}p_{\mu}$
and identify $w^{\mu}$ with the frame velocity. Such a choice is
equivalent to setting
\begin{equation}
A^{\mu\nu}=\beta(w^{\mu}p^{\nu}-w^{\nu}p^{\mu})\Omega^{-1}H(\Omega)
\end{equation}
in eqs.(38)-(39). Choosing a special frame $w=(1,0,0,0)$ gives
$A_{0j}=\beta p_{j}H(\Omega)\Omega^{-1}$ and $A_{jk}=0$ in
eq.(39). Assuming that the equilibrium $\Omega_{E}$ (40) depends
only on the energy (28)  it is easy to see that there exists an
equilibrium solution of eq.(40) if
\begin{equation}
\partial_{\epsilon}\Omega_{E}+H(\Omega_{E})=0.
\end{equation}
Then, $H(\Omega)=\Omega$ gives the J\"uttner distribution
\cite{juttner} whereas
\begin{equation}
H_{\sigma}(\Omega)=\Omega(1+\sigma\Omega)
\end{equation}
leads to the quantum distributions ($\sigma=+1$ for bosons,
$\sigma=-1$ for fermions and the Boltzmann statistics could be
treated as a limit $\sigma\rightarrow 0$) .
 In
general, the diffusion strength $G$ could again be a function of
$\Omega$.

 In a rest frame, the diffusion equation
(39) for scalar particles has the simple form
\begin{equation}\begin{array}{l}
\partial_{0}\Omega-p_{0}^{-1}{\bf p}\nabla_{{\bf x}} \Omega=\frac{1}{2}
\kappa^{2}\partial_{j}\Big(G(\Omega)\Big(
g^{jk}p_{0}^{-1}\partial_{k}\Omega  +\beta
cp^{j}H(\Omega)\Big)\Big).\end{array}
\end{equation}
In an arbitrary frame eq.(45) can be written in a covariant
divergence form (25)
\begin{equation}\begin{array}{l}
\sqrt{g}p^{\mu}\partial_{\mu}^{x}\Omega=
 \frac{1}{2}\kappa^{2}\partial_{j}\Big(GH g^{jk}\sqrt{g}
\Big(H^{-1}\partial_{k}\Omega +\exp(-\beta
cp_{\mu}w^{\mu})\partial_{k}\exp(\beta cp_{\mu}w^{\mu})
\Big)\Big).\end{array}
\end{equation}
If $\partial_{\mu}^{x}\Omega_{S} =0 $ in eq.(45) and $J\neq 0$
then $\Omega_{S}$ is called a stationary state. If
$J(\Omega_{E})=0$ then $\Omega_{E}$ is the equilibrium. Eq.(46)
has the equilibrium solution (this covariant form of the
equilibrium probability distribution is discussed in particle
physics in \cite{zeit}, see also \cite{temp})
\begin{equation}
\Omega_{E}=\Big(z\exp(\beta cw^{\mu}p_{\mu})-\sigma\Big)^{-1}
\end{equation} for  $H$ defined in eq.(44)( here $z=\exp(-\mu)$ and $\mu$ is the chemical
potential; the J\"uttner distribution corresponds to $\sigma=0$).

The equation with the drift $H_{q}(\Omega)=\Omega^{q}$
\begin{equation}\begin{array}{l}
\partial_{0}\Omega-p_{0}^{-1}{\bf p}\nabla_{{\bf x}} \Omega=
\frac{1}{2} \kappa^{2}\partial_{j}G(\Omega)\Big(
g^{jk}p_{0}^{-1}\partial_{k}\Omega  +\beta
cp_{j}\Omega^{q}\Big)\end{array}
\end{equation}
gives the Tsallis equilibrium distribution (34).

We would like to find a quantum  analog of the Tsallis distribution.
There are  already some candidates in the literature
\cite{quantumts}. We suggest another one resulting from the drift
\begin{equation}
H_{\sigma
 q}=\Big(\Omega(1+\sigma\Omega)\Big)^{q}\Big((1+\sigma\Omega)^{q}-\sigma\Omega^{q}\Big)^{-1}
\end{equation}
Eq.(43) gives an implicit equation for the equilibrium
\begin{equation}
\Omega_{E}^{1-q}-(1+\sigma\Omega_{E})^{1-q}=(q-1)(\epsilon-\mu).
\end{equation}
Eqs.(49)-(50) have the correct limit $q\rightarrow 1$ (quantum
distribution) and $\sigma\rightarrow 0$ (classical distribution).
We show in sec.7 that the drift $H_{\sigma q}$ allows to define a
monotonic relative entropy.

If we assume that $\Omega$ depends only on $\epsilon$ then eq.(45)
takes the simple form
\begin{equation}\begin{array}{l}
\partial_{t}\Omega_{t}(\epsilon)=\frac{\beta
c^{2}\kappa^{2}}{2}\epsilon^{-2}\Big(\partial_{\epsilon}\Big((\epsilon^{3}-
m^{2}c^{4}\beta^{2}\epsilon)\partial_{\epsilon}{\cal
G}(\Omega_{t})\Big)+\cr m^{2}c^{4}\beta^{2}\partial_{\epsilon}{\cal
G}(\Omega_{t}) +m^{2}c^{4}\beta^{2}GH+
\partial_{\epsilon}\Big(
(\epsilon^{3}-m^{2}c^{4}\beta^{2}\epsilon)GH\Big)\Big)
\end{array}\end{equation}
The stationary distribution is determined by the equation
$\partial_{t}\Omega=0$. After an integration of the rhs of eq.(51)
we obtain the first order equation for $\Omega_{S}$
\begin{equation}
G(\Omega_{S})(\frac{d\Omega_{S}}{d\epsilon}+H(\Omega_{S}))=R(\epsilon^{2}-m^{2}c^{4}\beta^{2})^{-\frac{3}{2}}
\end{equation}
where $R$ is an arbitrary integration constant. $R=0$ gives the
equilibrium (43). If $H=\Omega$ (corresponding to the J\"uttner
distribution ) and $G=1$ then eq.(52) is linear. It can be checked
that the only normalizable ($\int \Omega_{S}<\infty$) solution is
at $R=0$. However, there may be  normalizable solutions of eq.(52)
with $R\neq 0$ if $G\neq const$. The obvious example is $G\simeq
H^{-1}$.

In the limit $m\rightarrow 0$ eq.(45) takes the form resembling
the Kompaneets equation \cite{komp}
\begin{equation}
\partial_{t}\Omega-c{\bf n}\nabla_{{\bf
x}}\Omega=\frac{\beta
c^{2}\kappa^{2}}{2}\epsilon^{-2}\frac{\partial}{\partial
\epsilon}\Big(\epsilon^{3}G(\Omega)\Big(\frac{\partial}{\partial
\epsilon}\Omega+H(\Omega)\Big)\Big).
\end{equation}where ${\bf n}={\bf p}\vert{\bf p}\vert^{-1}$.

The stationary distribution is determined by
\begin{equation}
\frac{\partial}{\partial
\epsilon}\Omega_{S}+H(\Omega_{S})=R\epsilon^{-3}G(\Omega_{S})^{-1}.
\end{equation}
The Kompaneets equation has been derived in quantum electrodynamics
of electron-photon scattering \cite{komp}\cite{rybicki}\cite{zeld}
with a non-relativistic approximation for the electron velocity
distribution. In such a case a factor $\epsilon^{4}$ enters eq.(53)
instead of $\epsilon^{3}$. In refs.\cite{cooper}\cite{liv}
$\epsilon^{4}$ is replaced by a general function $\alpha(\epsilon)$
in order to take relativistic corrections into account. Our
discussion based on eqs.(14)(39)and (46) shows that the complete
differential equation in the phase space is relativistic covariant
only when $\alpha(\epsilon)=\epsilon^{3}$. The Kompaneets equation
has important applications in astrophysics and cosmology
(Sunyaeev-Zeldovitch effect). It is also studied as a mathematical
model for the time dependence of the Bose-Einstein condensation
\cite{zeld}\cite{pomeau}\cite{esp}.
\section{Low density approximation (linear diffusion)} We recall
the linear diffusion of ref.\cite{haba} and its thermodynamics
\cite{habampa} in order to compare it to the non-linear one. If we
assume that $\Omega$ is small then the general drift $H$ can be
expanded in a Taylor series
\begin{displaymath} H(\Omega)=\Omega+a_{2}\Omega^{2}+...
\end{displaymath} ($H(0)=0$ if the solution of eq.(43)
is to have the required property $\Omega_{E}\rightarrow 0$ for
$\epsilon\rightarrow \infty$ ). We assume in this section that
$G(\Omega)\simeq 1$ and $\Omega>>\Omega^{2}$. Then, eq.(45) becomes
linear
\begin{equation}\begin{array}{l}
\partial_{0}\Omega-p_{0}^{-1}{\bf p}\nabla_{{\bf x}} \Omega=
\frac{1}{2}\kappa^{2}\partial_{j}\Big(
g^{jk}p_{0}^{-1}\partial_{k}\Omega +\beta
cp_{j}\Omega\Big).\end{array}
\end{equation} It has
 the J\"uttner equilibrium solution \cite{juttner}.

We wish to describe the time evolution in terms of some extensive
(thermodynamic) functions of the probability density. It seems
that the relative entropy measuring the distance between two
probability distributions plays the crucial role in such a
description (the relative entropy measures the speed of
convergence $\Omega_{t}\rightarrow\Omega_{E}$, its relevance has
been discovered in \cite{lebo}\cite{haken}; the relative entropy
in nonextensive statistical mechanics is discussed in \cite{book})
 We define the Boltzmann-Kullback-Leibler relative entropy $S_{K}$ of two
unnormalized probability distributions $\Omega$ and $\Omega_{E}$
($N$ and $N_{E}$ are the normalization constants)
\begin{equation}
S_{K}(\Omega;\Omega_{E})=N^{-1}\int d{\bf x}d{\bf p}\Omega\ln\Big(
N^{-1}\Omega(\Omega_{E})^{-1}N_{E}\Big).\end{equation}  It is
known that \cite{risken}
\begin{equation} S_{K}(\Omega;\Omega_{E})\geq 0.
\end{equation}
An easy calculation using the diffusion equation (55) gives
\begin{equation}\begin{array}{l}
\partial_{0}S_{K}(\Omega;\Omega_{E})
=-\frac{1}{2}N^{-1}\int d{\bf p}d{\bf x}p_{0}^{-1}\Omega^{-1}
g^{jk}(\partial_{j}\Omega+\beta
cp_{j}p_{0}^{-1}\Omega)(\partial_{k}\Omega+\beta
cp_{k}p_{0}^{-1}\Omega).
\end{array}\end{equation}
 It follows from eqs.(57)-(58) that
$S_{K}(\Omega;\Omega_{E})$ is a non-negative function
monotonically decreasing to zero  when $t\rightarrow \infty$.

We  define the Boltzmann entropy
\begin{equation}
S(\Omega)=-k_{B}\int d{\bf x}d{\bf p}\Omega\ln\Omega
\end{equation} the  energy
\begin{equation}
{\cal E}=c\int d{\bf p}d{\bf x}p_{0}\Omega
\end{equation}
and the free energy
\begin{equation} {\cal
F}=\beta^{-1}N\Big(S_{K}(\Omega;\Omega_{E})-\ln(NN_{E}^{-1})\Big).
\end{equation}
Then, from eqs.(58)-(61)  we obtain the basic thermodynamic
relation
\begin{equation}TS={\cal E}-{\cal F}.
\end{equation}

\section{Thermodynamics of the non-linear diffusion}
In general, a meaningful (non-Hamiltonian) evolution equation
should lead to an increase of entropy. Conversely, if the entropy
is defined then the evolution equation could be determined from
the requirements that the entropy is increasing and bounded. If we
define the entropy $S$ in terms of an entropy density $s$
\begin{equation}
S=-\int d{\bf x}d{\bf p}s(\Omega)
\end{equation}
then for the diffusion (22) (without drift)
\begin{equation}
\partial_{0} S =\frac{1}{2}\int d{\bf x}d{\bf p}p_{0}^{-1}
s^{\prime\prime}(\Omega)G(\Omega)\partial_{j} \Omega
g^{jk}\partial_{k} \Omega .\end{equation} It follows that $S$ is
increasing if
\begin{equation}
s^{\prime\prime}\geq 0.
\end{equation}
The diffusion without a drift has no equilibrium but only a
stationary state. There may be no limit for the increase of the
entropy. For diffusions with the drift (44) quadratic in $\Omega$
(quantum equilibrium distributions) as well as for
$H_{q}=\Omega^{q}$ and $H_{\sigma q}$ (49) (Tsallis distributions)
we can define the relative entropy $S_{K}(\Omega,\Omega_{E})$
which is monotonically decreasing to the equilibrium state when
$t\rightarrow \infty$ ($S_{K}$ of eq.(58) would not be a monotonic
function). For the drift (44) define
\begin{equation}\begin{array}{l} S_{K}^{\sigma}(\Omega,\Omega_{E})=\int
d{\bf x}d{\bf
p}\Big(\Omega\ln(\Omega\Omega_{E}^{-1})-\sigma(1+\sigma\Omega)\ln\Big(
(1+\sigma\Omega)(1+\sigma\Omega_{E})^{-1}\Big)\Big)\end{array}
\end{equation}
 for $H=\Omega^{q}$ (such a relative entropy is suggested also in the Appendix of
 \cite{rajagopal}; for a general  theory see \cite{book})
 \begin{equation}
\begin{array}{l}S_{K}^{q}(\Omega,\Omega_{E})= (2-q)^{-1}(1-q)^{-1}\int d{\bf
x}d{\bf p}\Omega^{2-q}\cr -(1-q)^{-1}\int d{\bf x}d{\bf
p}\Omega\Omega_{E}^{1-q}+(2-q)^{-1}\int d{\bf x}d{\bf
p}\Omega_{E}^{2-q}.\end{array}
\end{equation}We define Tsallis relative quantum entropy
for the drift  $H_{\sigma q}$ (49) \begin{equation}
\begin{array}{l}S_{K}^{\sigma q}(\Omega,\Omega_{E})= (2-q)^{-1}(1-q)^{-1}\int d{\bf
x}d{\bf p}\Omega^{2-q}\cr -(1-q)^{-1}\int d{\bf x}d{\bf
p}\Omega\Omega_{E}^{1-q}+(2-q)^{-1}\int d{\bf x}d{\bf
p}\Omega_{E}^{2-q}\cr -\sigma(2-q)^{-1}(1-q)^{-1}\int d{\bf
x}d{\bf p}(1+\sigma\Omega)^{2-q} +\sigma(1-q)^{-1}\int d{\bf
x}d{\bf p}(1+\sigma\Omega)(1+\sigma\Omega_{E})^{1-q}\cr
-\sigma(2-q)^{-1}\int d{\bf x}d{\bf
p}(1+\sigma\Omega_{E})^{2-q}.\end{array}
\end{equation}
The three entropies (66)-(68) have the properties $
S_{K}(\Omega_{E},\Omega_{E})=0$ , $
S_{K}^{\prime}(\Omega_{E},\Omega_{E})=0$ and $
S_{K}^{\prime\prime}(\Omega,\Omega_{E})>0$. It follows that
$S_{K}(\Omega,\Omega_{E})> 0$(in contradistinction to $S_{K}$ in
eq.(58) we do not need to normalize $\Omega$ and $\Omega_{E}$ in
order to prove the positivity of $ S_{K}$ in (66)-(68))  .

In an equilibrium state
\begin{equation}
J^{r}=G(\Omega_{E}) g^{rk}p_{0}^{-1}\Big(\partial_{k}\Omega_{E}
+\beta cp_{k}p_{0}^{-1}H(\Omega_{E})\Big)=0.
\end{equation}
Then, using eq.(39) for the calculation of  the time derivative of
$S_{K}$ we obtain \begin{equation}\begin{array}{l}
\partial_{0} S_{K}(\Omega,\Omega_{E})=
-\frac{1}{2}\int d{\bf x}d{\bf p}p_{0}^{-1}H(\Omega)^{-1}
G(\Omega)g^{jk}(\partial_{j} \Omega
+A_{j\mu}p^{\mu}\Omega)(\partial_{k} \Omega
+A_{k\mu}p^{\mu}\Omega) )\end{array}\end{equation} (the Lorentz
invariance of eq.(70) follows from eq.(14)). In the rest frame
\begin{equation}\begin{array}{l}
\partial_{0} S_{K}(\Omega,\Omega_{E})=
-\frac{1}{2}\int d{\bf x}d{\bf p}p_{0}^{-1}H(\Omega)^{-1}
G(\Omega)\cr
g^{jk}(\partial_{j} \Omega +\beta
cp_{j}p_{0}^{-1}H(\Omega))(\partial_{k} \Omega +\beta
cp_{k}p_{0}^{-1}H(\Omega)) )\end{array}\end{equation}
 where $H$ is
either $H_{\sigma}$,$H_{q}$ or $H_{\sigma q}$.

In the massless case eq.(71) simplifies
to\begin{equation}\begin{array}{l}
\partial_{0} S_{K}(\Omega,\Omega_{E})=-\frac{1}{2}\int d{\bf p}p_{0}^{-1}H(\Omega)^{-1}
G(\Omega)(p^{j}\partial_{j} \Omega +\beta
cp_{0}H(\Omega))^{2}.\end{array}\end{equation} It follows that  $
S_{K}(\Omega,\Omega_{E})>0$ and the time evolution will decrease $
S_{K}$ until it achieves its minimum at the equilibrium (then
$S_{K}(\Omega_{E},\Omega_{E})=0$). If the state depends solely on
the energy then ( this equality has been derived in the particular
case of the Kompaneets equation in ref.\cite{liv})
\begin{equation}
\partial_{0}S_{K}(\Omega,\Omega_{E})=-4\pi\int d\epsilon
\epsilon^{3}GH^{-1}(\partial_{\epsilon}\Omega+H)^{2}.
\end{equation}
In the stationary state $\Omega_{S}$ (54) (when $J\neq 0$) there
is a correction to the
 formulas (70)-(73). So, in the massless case under the assumption
 that both $\Omega$ and $\Omega_{S}$ depend only on $\epsilon$ we
 obtain\begin{equation}\begin{array}{l}
 \partial_{0}S_{K}(\Omega,\Omega_{S})=-4\pi\int d\epsilon
 \epsilon^{3}G(\Omega)H(\Omega)^{-1}(\partial_{\epsilon}\Omega+H(\Omega))^{2}
\cr +4\pi R\int d\epsilon
G(\Omega)G(\Omega_{S})^{-1}H(\Omega_{S})^{-1}(\partial_{\epsilon}\Omega+H(\Omega)).
\end{array} \end{equation} The $R$-correction has an indefinite
sign. Eq.(74) shows that $S_{K}(\Omega,\Omega_{S})$ is not
monotonic. $\Omega_{t}$ does not tend to $\Omega_{S}$. It is
convergent to $\Omega_{E}$.

 We define the entropy (as suggested by eqs.(66)-(68)) for the
Bose-Einstein and Fermi-Dirac distributions
\begin{equation} {\cal S}_{\sigma}=-k_{B}\int d{\bf x}d{\bf
p}\Big(\Omega\ln\Omega-\sigma(1+\sigma\Omega)\ln
(1+\sigma\Omega)\Big)
\end{equation}
and for the quantum Tsallis distribution
\begin{equation}\begin{array}{l} {\cal S}_{\sigma q}=-k_{B}
(2-q)^{-1}(1-q)^{-1}\int d{\bf x}d{\bf p}\Omega^{2-q}
\cr
+k_{B}\sigma(2-q)^{-1}(1-q)^{-1}\int d{\bf x}d{\bf
p}(1+\sigma\Omega)^{2-q} \end{array}\end{equation} (when $\sigma=0$
we have ${\cal S}_{\sigma q}\rightarrow {\cal S}_{q}$). In the
models discussed in this section the principle of the maximum of
entropy is satisfied. $S$ is achieving the maximum (with fixed $\int
\Omega=1$ and $\int \epsilon\Omega={\cal E}$) at the equilibrium
state $\Omega_{E}$. We define the free energy as ${\cal F}={\cal
E}-T{\cal S}$ ( where ${\cal E}$ is defined in eq.(60)). From
eqs.(66)-(68) and (75)-(76) we obtain that ${\cal F}=S_{K}+{\cal
F}_{E}$ where ${\cal F}_{E}$ is the time independent equilibrium
free energy. Then, eq.(62) is satisfied as an identity.

\section{Discussion and summary}
The non-linear diffusion equations result from a macroscopic
averaging of complex microscopic phenomena. We have shown that the
basic requirements imposed on such equations determine their form.
The linear part of the equation is already determined by the
Lorentz invariance and the requirement of the positivity of the
probability density. The  non-linearity of the equation comes from
quantum statistics (dependence of the transition rate to a state
on the occupation of the state) and  from an elimination of some
fields which are averaged in the reduced dynamics. We have shown
that the form of the non-linearity is closely related with the
formula for the entropy. It is possible to obtain a drift for a
linear diffusion such that the diffusion is equilibrating to the
Bose-Einstein or Fermi-Dirac equilibrium distribution \cite{haba}.
However, it seems that the non-linear drift is indispensable if we
wish to define the relative entropy as a monotonically decreasing
function which  satisfies the thermodynamic relation (62). We can
apply the non-linear diffusion equation to a statistical
description of the motion of a stream of relativistic particles in
a medium of some other relativistic particles. Such a problem is
studied in high-energy physics \cite{groot} and astrophysics
\cite{bernstein}\cite{silk}. However, the whole space-time
dependence of the phase-space distribution is not discussed in
these papers. It can be a complicated problem to derive an
equation of space-time evolution of the stream of particles on the
basis of a theory of fundamental interactions (see
\cite{rybicki}\cite{elze}\cite{muller}\cite{kleinert}). It may
require non-perturbative methods. In general, the resulting
equation will be non-local. If the Markovian approximation is
applicable to the full space-time dynamics then the relativistic
invariance leads to the equations discussed in this paper. The
result of the multi-particle ultra-relativistic scattering may be
independent of the details of the interaction but depend more on
kinematics and the
 relativistic processes  of dissipation. In such a case we could
 check in experiments  assumptions underlying the dissipative
 dynamics of eq.(6).
 Diffusion equations already served as
a theoretical basis for RHIC data analysis \cite{ion}\cite{wol}.
The diffusion model could be tested in heavy ion collision
experiments. The complete space-time diffusion equation is also
interesting as a model for space-time evolution of the
Bose-Einstein condensation (discussed in atomic physics
\cite{pomeau} and in astrophysics as a model for the star
formation from the dark matter \cite{dark}).

\end{document}